\documentclass[12pt,preprint]{aastex}

\newcommand{\ud}{\mathrm{d}}
\newcommand{\ue}{\mathrm{e}}
\newcommand{\vp}{\varphi}
\newcommand{\ra}{\rightarrow}
\newcommand{\be}{\begin{equation}}
\newcommand{\ee}{\end{equation}}
\newcommand{\bea}{\begin{eqnarray}}
\newcommand{\eea}{\end{eqnarray}}

\newcommand{\wt}{\widetilde}
\DeclareGraphicsExtensions{.eps}

\begin{document}

\title{
Spontaneous violation of the energy conditions}

\author{A. W. Whinnett\footnote{Astronomy Centre, University of Sussex, Falmer,
Brighton, BN1 9QJ,  UK. E-mail: visitor2@pact.cpes.susx.ac.uk}, \&
Diego F. Torres\footnote{Lawrence Livermore National Laboratory,
7000 East Ave., L-413, Livermore, CA 94550. E-mail:
dtorres@igpp.ucllnl.org}}

\begin{abstract}

A decade ago, it was shown that a wide class of scalar-tensor
theories can pass very restrictive weak field tests of gravity and
yet exhibit non-perturbative strong field deviations away from
General Relativity. This phenomenon was called `Spontaneous
Scalarization' and causes the (Einstein frame) scalar field inside
a neutron star to rapidly become inhomogeneous once the star's
mass increases above some critical value. For a star whose mass is
below the threshold, the field is instead nearly uniform (a state
which minimises the star's energy) and the configuration is
similar to the General Relativity one. Here, we show that the
spontaneous scalarization phenomenon is linked to another strong
field effect: a spontaneous violation of the weak energy
condition.
\end{abstract}


\section{Introduction}

Scalar-tensor (ST) theories  describe gravity as being mediated by
both a metric $g_{ab}$ and a scalar field $\Phi$. The latter is
coupled to the metric via a function $\omega(\Phi)$. These
theories are fully conservative and only two parameterised
post-Newtonian (PPN) parameters, $\gamma$ and $\beta$, appear in
the formalism. Currently, they are constrained to have the values
$|\gamma-1|\leq 0.0003$ and $|\beta-1|\leq 0.002$ (see Will 1998
for details). For Brans-Dicke-like (BD) theories, the first of
these inequalities implies that its coupling is today $\omega >
3300$. All predictions of BD theory differ from those of General
Relativity (GR) to within a relative deviation of $\sim 1/\omega$.
For other, more general, ST theories, the observational data only
place limits on the behaviour of $\omega(\Phi)$ in the slow
motion, weak field limit. However, strong field tests generally
place weaker constraints on ST theories than those given above.
Hence, it is possible to have a ST theory which satisfies all
current constraints but shows strong field effects that are
significantly different from GR. In this Letter we present the
case for the appearance of a spontaneous violation of the energy
conditions as one of such strong field effects.

\section{Scalarization}

Spontaneous scalarization was discovered by Damour \&
Esposito-Farese (1993, 1997) in scalar-tensor models of neutron
stars. They found that for particular forms of the function
$\omega(\Phi)$, see the Lagrangian density below, the $\Phi$ field
inside a neutron star rapidly becomes inhomogeneous once the
star's mass increases above some critical value. For a star whose
mass is below this value, $\Phi$ is nearly homogeneous throughout
the star (a state which minimises the star's energy), while for
higher mass stars, the energy is minimised when the field has a
large spatial variation. To exhibit these effects, $\omega(\Phi)$
must be such that its derivative with respect to $\Phi$ satisfies
the inequality ${\wt\beta}_{0}:=2\Phi_{B}(2\omega+3)^{-2}\;
\ud\omega/\ud\Phi_{B} < -4$, where the subscript `{\scriptsize B}'
denotes cosmological or `background' quantities evaluated far from
strongly gravitating sources. Theories in which $\omega(\Phi)$
satisfies the above inequality may be arbitrarily close to GR in
the weak field limit but yet significantly diverge from it in
strong field regions. The scalarization effect becomes more
pronounced as $\Phi_{B}\ra 1$.

The Jordan frame (JF) action for the ST theories  is
\be\label{JFaction} I=\int \ud^{4}x\sqrt{-g}\left(R\Phi
-\frac{\omega}{\Phi}g^{ab}\nabla_{a}\Phi\nabla_{b}\Phi +16\pi
L_{m}\right), \ee where $R$ is the Ricci scalar and $L_{m}$ is the
matter Lagrangian. The field equations in the JF are
$
G_{ab}=S_{ab}+{8\pi}T_{ab}/{\Phi},
$ where $T_{ab}$ is the matter energy-momentum tensor, $G_{ab}$ is
the Einstein tensor,  and $S_{ab}$ is given by \bea\label{Sabdef}
S_{ab}=\frac{1}{\Phi}(\nabla_{a}\nabla_{b}\Phi
   -g_{ab}\nabla^{c}   \nabla_{c}\Phi)+
\frac{\omega}{\Phi^{2}}  \nonumber \\
(\nabla_{a}\Phi\nabla_{b}\Phi -\frac 12
g_{ab}g^{cd}\nabla_{c}\Phi\nabla_{d}\Phi). \eea

To rewrite the action and the resulting field equations in the
Einstein frame (EF), in which the metric is ${\wt g}_{ab}$, one
makes the field redefinition
$\ud\vp=\ud\Phi{\sqrt{2\omega+3}}/({2\Phi})$, which determines the
relationship between $\omega$ and $\vp$. One also needs to define
the relationship between $\Phi$ and $\vp$, and we adopt the
notation ${\cal A}(\vp):={1}/{\sqrt{\Phi}}$. The conformal
transformation between the EF and JF metric is then ${\wt
g}_{ab}={\cal A}^{-2}g_{ab}$. In the EF, the ST theory is
determined by the functional form of ${\cal A}(\vp)$.

We are primarily concerned with analysing the internal structure
of static, spherically symmetric neutron star solutions in the JF
and, in particular, looking for violations of the weak energy
condition (WEC) in this frame. However,  it is easier to solve the
field equations in the EF when scalarization effects are present.
In addition, the general vacuum spherically symmetric ST solution
is known in the EF (Coquereaux \& Esposito-Farese 1990) and we
need to match our solutions to a vacuum exterior, both to fix the
value of the central scalar field and to find the solutions'
masses. Hence we solve the EF equations of structure here, details
of which can be found in the work by Damour and Esposito-Farese
(1993).

To characterise the degree to which the scalar field is
inhomogeneous, a convenient parameter is the scalar charge $Q_{S}$
which, in the Jordan frame and for a spherically symmetric
solution, is defined to be \be
Q_{S}:=\lim_{r\to\infty}\left(r^{2}\frac{\ud \Phi}{\ud r}\right),
\ee where $r$ is the Schwarzschild radial coordinate. This
quantity is used to evaluate the conserved JF energy of the star,
\be\label{MTdef}
  M_{T}:=M_{ADM}-\frac 12 Q_{S}
\ee  where $M_{ADM}$ is the ADM mass in the JF (Lee 1974). In the
EF, the scalar charge ${\wt Q}_{S}$ associated with $\vp$ is
defined in a analogous way. The EF scalar charge may be used to
define an effective coupling strength $\alpha:={\wt Q}_{S}/{\wt
M}_{ADM}$, where ${\wt M}_{ADM}$ is the ADM mass in the EF. This
latter quantity is identical to the JF quantity $M_{T}$ up to a
conformal factor which is close to unity. In terms of JF
quantities, the effective coupling $\alpha$ may be rewritten as
$\alpha=(1/2) \sqrt{2\omega_{B}+3}\;Q_{S}/M_{T}$. Hence the EF
quantity $\alpha$ is also a good indicator of scalar field
inhomogeneity in the JF.

\section{Energy conditions}

In GR, the strong, dominant, and weak energy conditions are
usually formulated by placing restrictions on the properties of
the matter in a solution (see, for example, Hawking \& Ellis
1973). By virtue of the GR field equations $G_{ab}=8\pi T_{ab}$,
any condition placed on $T_{ab}$ is automatically satisfied by
$G_{ab}$. Thus, in GR, the energy conditions are also statements
about the geometry of any given solution. The singularity
theorems, for example, rely on the behaviour of $G_{ab}$ and make
no explicit reference to the matter content of the theory. In ST
gravity, the situation is a little more complex. The field
equations allow both $G_{ab}$ and $T_{ab}$ to obey different
conditions in the same solution. In particular, in a solution in
which the normal matter obeys all three energy conditions,
$G_{ab}$ need not obey any. This is because $S_{ab}$, defined by
Eq. (\ref{Sabdef}), contains terms that are linear in the second
derivatives of $\Phi$, and no restriction on the sign of these
derivatives apply (see  Soko{\l}owski 1989a,b; Cho 1992; Magnano
\& Soko{\l}owski 1994, Torres 2002 for discussions). Hence one can
have a solution in which the normal, bosonic or fermionic, matter
has a positive energy density and yet the WEC is violated. This
would have implications for, for example, singularity theorems and
other statements about the allowed global structure of a spacetime
(see Anchordoqui et al. 1996 for an example, and Barcelo \& Visser
2000 for further discussion). Hence it is reasonable to
investigate whether spontaneous scalarization is accompanied by a
spontaneous violation of WEC.

The WEC holds when the quantity $G_{ab}K^{a}K^{b}$ is non-negative
for all timelike and null vectors $K^{a}$. In the JF, a local
effective energy density $8\pi\mu:=G_{ab}U^{a}U^{b}$ may be
defined, where $U^{a}$ is a unit timelike vector. This quantity is
not a true energy density, since it mixes up contributions from
both the matter and the $\Phi$ field. However, it is a useful
quantity since a sufficient condition for WEC violation is
$\mu<0$. Since we are concerned here with static solutions, we
shall take $U^{a}$ to be the timelike unit Killing vector field.
In terms of the metric and scalar fields, using the wave equation
for $\Phi$ one can show that \bea\label{Density}
   8\pi\mu=\frac{1}{\Phi}U^{a}U^{b}\nabla_{a}\nabla_{b}\Phi
   +\frac{\omega}{2\Phi^{2}}\;g^{ab}\nabla_{a}\Phi\nabla_{b}\Phi
   \;\;\;\;\;\; \nonumber \\
   +\frac{1}{\Phi}\left(8\pi\rho+\frac{8\pi (3p-\rho)}{2\omega+3}
   -\frac{1}{2\omega+3}\nabla^{a}\Phi\nabla_{a}\Phi\;
   \frac{\ud\omega}{\ud\Phi}\right),
\eea where we have used the assumption that $T_{ab}$ is that of a
perfect fluid. Integrating this quantity over a spacelike
hypersurface orthogonal to $U^{a}$ gives the ADM mass in the JF.

\section{Neutron star solutions in ST gravity}

As an example, we shall examine neutron star solutions in the two
ST theories considered by Damour \& Esposito-Farese (1993). The
first theory is specified by the exponential function ${\cal
A}(\vp)=\ue^{-\kappa\vp^{2}}$, for which
${\wt\beta}_{0}=-2\kappa$. This is equivalent to the JF theory
with $2\omega+3={1}/({2\kappa\log\Phi})$. The second theory we
consider is specified by the function ${\cal
A}(\vp)=\cos(\sqrt{\lambda}\,\vp)$, for which
${\wt\beta}_{0}=-\lambda$. This is equivalent to the JF theory
with $2\omega+3=1/[\lambda(\Phi-1)]$. For both of these theories,
the PPN constraint on $\gamma$ is the most restrictive.

The above forms of the function $\omega(\Phi)$ diverge as
$\Phi_{B}$ approaches the GR limit. For this reason, the JF
descriptions of these theories break down and it is more
appropriate to use field variables of a JF formalism developed by
Damour \& Esposito-Farese (1992, 1996) and in which the results of
Damour \& Esposito-Farese  (1993, 1997)  were presented.

We have integrated the equations of structure in the EF for
neutron stars with the polytropic equation of state (EOS) \be
\rho=nm+\frac{Kn_{0}m}{\Gamma-1}\left(\frac{n}{n_{0}}\right)^{\Gamma},
\;\;\;\;p=Kn_{0}m\left(\frac{n}{n_{0}}\right)^{\Gamma}, \ee where
$m$ is the neutron mass, $n_{0}=1.0\times 10^{44}\,$m$^{-3}$ and
the parameters $K$ and $\Gamma$ have the values $K=0.0195$,
$\Gamma=2.34$. This EOS was used by Damour \& Esposito-Farese
(1993) as a best fit to a realistic description of high density
neutron matter given by Diaz-Alonso \& Ibanez-Cabanell (1985). We
use the maximum values of $\vp_{B}$ consistent with the above PPN
constraints (which are tighter than those used by Damour \&
Esposito-Farese 1993). For each solution, we compute effective
density profiles in the JF, using Equation (\ref{Density}).

\begin{figure*}[t]
\begin{center}
\includegraphics[width=0.8\textwidth]{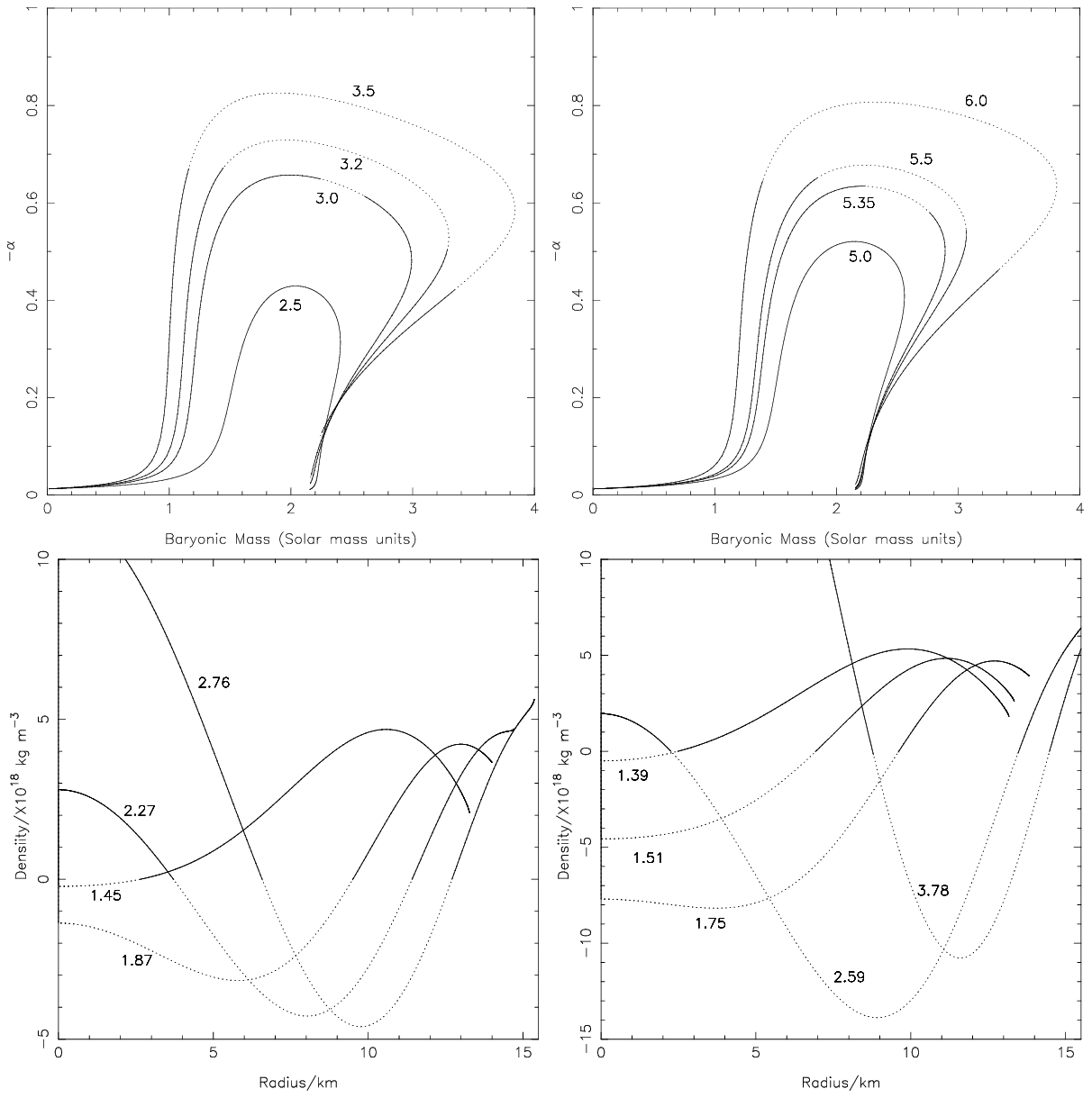}
\caption{\label{stars} Top: Mass against scalar coupling for
neutrons stars in the theory with ${\cal
A}(\phi)=\ue^{-\kappa\vp^{2}}$ (left) and ${\cal
A}(\phi)=\cos(\sqrt{\lambda}\,\vp)$ (right). Bottom: Density
against radius for neutron stars in theory with ${\cal
A}(\phi)=\ue^{-\kappa\vp^{2}}$ (left) and ${\cal
A}(\phi)=\cos(\sqrt{\lambda}\,\vp)$ (right).}
\end{center}
\end{figure*}

Fig.~\ref{stars} (top panels) shows curves of $\alpha$ against
baryonic mass $M$ for solutions in both ST theories described
above. The left panel is for the exponential theory ${\cal
A}(\vp)=\ue^{-\kappa\vp^{2}}$, while the right is for the cosine
theory ${\cal A}(\vp)=\cos(\sqrt{\lambda}\,\vp)$. Aside from the
lower values of $\vp_{B}$, the curves are similar to those shown
by Damour and Esposito-Farese (1993). However, they show an
additional feature not reported in previous work in that some of
the solutions, denoted by the dotted segments on some of the
curves, contain a region in which the effective density $\mu$ is
negative. Hence, spontaneous scalarization appears to be linked
with a second phenomenon, that of spontaneous WEC violation.

As Fig.~\ref{stars} (top panels)  shows, only large mass stars
which are already showing the scalarization effect exhibit WEC
violation. The reason for this is as follows. From Equation
(\ref{MTdef}), the energy of a star depends upon both $M_{ADM}$
(which can be written as an integral of $\mu$) and $Q_{S}$. This
latter quantity is negative for all solutions, so tends to
increase the star's energy. For a weak field star, $\vp$ and hence
$\Phi$ are nearly homogeneous, $Q_{S}$ is vanishingly small and
$\mu$ is dominated by the matter density $\rho$. At the onset of
scalarization, the increase in $M_{T}$ due to a non-zero $Q_{S}$
is cancelled by the decrease in $M_{ADM}$ caused by the non-zero
gradients in $\Phi$ reducing $\mu$ over much of the star's
interior. Once this happens, an inhomogeneous $\Phi$ becomes
energetically favoured and in some cases this causes $\mu$ to be
negative in some regions of the star.

The existence of WEC violation also depends strongly on the chosen
value of ${\wt\beta}_{0}$. For the exponential theory, WEC
violation occurs when $\kappa\ge 3.0$, corresponding to
${\wt\beta}_{0}< -6.0$. For the cosine theory, WEC violation
occurs when $\lambda\ge 5.35$, corresponding to ${\wt\beta}_{0}\le
-5.35$. It is apparent from these results that, for the same
constraints on $\vp_{B}$, different theories can exhibit WEC
violation at different values of ${\wt\beta}_{0}$ and there is no
reason to believe that other choices of ${\cal A}(\vp)$ would not
show negative energy density at larger values of ${\wt\beta}_{0}$.

Finally, we have found that larger values of $\vp_{B}$ decrease
the value of ${\wt\beta}_{0}$ at which WEC violation occurs. For
example, by increasing the value of $\vp_{B}$ by a factor of
approximately 15 (corresponding to an increase of $\Phi_{B}$ by
less than $10\%$), we have found that negative density a region
appears in stars in the exponential theory when $\kappa\ge 2$,
corresponding to ${\wt\beta}_{0}\le -4$.

Damour and Esposito-Farese (1997) have found that ST theories must
satisfy the constraint ${\wt\beta}_{0}>-5$. This bound is based on
a study of the binary pulsar in the exponential theory and is
approximate, so could conceivably be slightly lower than reported.
We have found that the cosine theory exhibits WEC violation at a
value of ${\wt\beta}_{0}$ which is greater than that for the
exponential coupling and closer to the inferred bound. There is no
reason why other ST theories might not show spontaneous WEC
violation at values of ${\wt\beta}_{0}>-5$. Even if the bounds on
${\it\beta}_{0}$ were to exclude WEC violating solutions in this
or other ST theories, one could still see this effect in real
neutron stars: we have assumed here that each star is isolated and
has a $\vp$ field which matches to the cosmological background
field $\vp_{B}$. However, a neutron star in the region of, for
example, a strongly gravitating companion would be subject to
different boundary conditions on its internal scalar field and, as
we have discussed, this may allow WEC violation at a larger value
of ${\wt\beta}_{0}$.

To show where within a neutron star WEC-violating regions occur,
we have computed density profiles for several of the solutions
shown in Fig.~\ref{stars} (top panels). The bottom panels of
Fig.~\ref{stars} show the JF energy density as a function of
Schwarzschild radius for several solutions containing WEC
violating regions. Each curve is labelled by its baryonic mass and
each curve terminates at the surface of the star. At this point,
the density is dominated by the scalar field and is always
positive. The dotted potion of each curve shows where the WEC is
violated. The plot on the left shows solutions in the exponential
theory and all are for $\kappa=3.2$. The WEC violating solutions
fall into two types. For the low mass solutions, the negative
energy density region begins at the centre and extends part of the
way out towards the surface. As the baryonic mass increases, the
star's central density eventually becomes positive and the WEC
violating region encloses a positive density core, as in the lower
$\kappa$ case. The primary reason for this is that, for larger
mass solutions, the central matter density is very large and
dominates over the WEC violating terms in $S_{ab}$. Only away from
the star's centre, when the matter density falls, can the negative
scalar field energy density start to dominate. The plot on the
right of Fig.~\ref{stars} (bottom panels) shows density plots for
the cosine theory and all are for $\lambda=6.0$. These are similar
to those for the exponential theory. Density profiles for other
values of ${\wt\beta}_{0}$ are similar.

\section{Concluding Remarks}

We have shown that spontaneous scalarization is accompanied by a
second effect, that of spontaneous weak energy condition
violation. This effect occurs for ST theories which give positive
effective energy densities for matter solutions in a weak field
region. It is only within a strong field regime, in compact
objects, that the effect occurs. It is important to note that this
is an effect limited in extent to inner regions of neutron stars.
A distant observer will not see a negative energy or negative
Kepler mass object (at least not according to our present, static,
simulations) so it would not be possible to use the techniques
outlined by Safonova et al. (2002) to observe these objects. There
certainly could be, however, other observable effects in the
evolution of the neutron star itself.

All in all, although we have shown that a mechanism for energy
conditions violation is linked to a plausible evolutionary process
(i.e. scalarization) in a known astrophysical environment (i.e.
neutron stars), the stability of the latter under a scalarization
process remains to be studied dynamically. It might be the case
that no neutron stars are stable if scalarization occurs (due to
the energy condition violation herein discovered); this would
imply strong constraints on the plausibility of scalar-tensor
theories to represent the observed Universe.

A natural extension of this work would be to determine whether and
how a star can evolve from a positive energy density state to one
which violates the WEC: our present study examines only static
solutions. Three possible routes seem likely. The first is through
accretion of matter. One can envisage a neutron star, whose
baryonic mass is just below that needed for WEC violation,
accreting matter to raise its mass to above the critical value.
Although we do not have details of the time evolution of this
process, it is reasonable to assume that the initial and final
states correspond to a static solution similar to those discussed
here. A second possibility is through gravitational collapse. For
the ST theories we consider here, one needs strong gravitational
fields to produce a $\Phi$ field that is inhomogeneous enough to
cause WEC violation. Hence the interior of a massive star just
before core collapse and neutron star formation would have a
constant scalar field and a positive energy density. Once the
neutron stars has been formed, the central density would then be
large enough for WEC violation to occur. A third possibility is
through gravitational evolution (as explored by, e.g., Torres
1997; Torres et al. 1998a,b; Whinnett \& Torres 1999 for boson
stars). In general, the background field $\vp_{B}$ (and its JF
equivalent $\Phi_{B}$) should increase as the Universe expands.
Assuming that
the star evolves adiabatically and can be modelled as a sequence
of static equilibrium solutions, there may come a point in the
star's lifetime when $\vp_{B}$ is large enough for it to suddenly
acquire a WEC violating region. This possibility will be explored
elsewhere.

\acknowledgements
We gratefully acknowledge Gilles Esposito-Farese and Israel Quiros
for a critical reading of the manuscript. AW thanks the Univerity
of Sussex for its hospitality while this work was completed. DFT
further acknowledges L. Anchordoqui for often discussions on this
topic. The  work of DFT was performed under the auspices of the
U.S. Department of Energy (NNSA) by University of California
Lawrence Livermore National Laboratory under contract No.
W-7405-Eng-48.


\end{document}